\documentclass[a4paper, 10pt, conference]{ieeeconf}      

\IEEEoverridecommandlockouts                              

\usepackage[hidelinks]{hyperref}
\usepackage{graphicx}
\usepackage{subcaption}
\usepackage[section]{placeins}
\usepackage{color}
\usepackage{adjustbox}
\usepackage{todonotes}

\usepackage[ruled,noend,noline]{algorithm2e}
\usepackage{placeins}
\usepackage{tikz}
\SetCommentSty{mycomment}
\DontPrintSemicolon
\SetKwInOut{Input}{input}
\SetKwInOut{Inout}{in/out}
\newenvironment{myalgorithm}[1][tbp] 
  {\begin{algorithm}[#1]\setlength{\parskip}{0.4ex}} 
  {\end{algorithm}}

\title{\LARGE \bf
Compressed Bounding Volume Hierarchies for Collision Detection \& Proximity Query
}

\author{Toni Tan$^{1}$, Ren\'{e} Weller$^{1}$ and Gabriel Zachmann$^{1}$
\thanks{$^{1}$University of Bremen, Germany%
        }
}

\begin{document}

\maketitle
\thispagestyle{empty}
\pagestyle{empty}

\begin{abstract}
We present a novel representation of compressed data structure for simultaneous bounding volume hierarchy (BVH) traversals like they appear for instance in collision detection \& proximity query. The main idea is to compress bounding volume (BV) descriptors and cluster BVH into a smaller parts (“treelet”) that fit into CPU cache while at the same time maintain random-access and automatic cache-aware data structure
layouts. To do that, we quantify BV and compress “treelet” using predictor-corrector scheme with the predictor at a specific node in the BVH based on the chain of BVs upwards.

\end{abstract}

\section{Introduction}

\emph{Collision detection (CD)} \& \emph{Proximity Query (PQ)} are essential for motion planning, especially sampling-based algorithms. They are used to test collision or calculate separation distance between sampled configuration and workspace obstacles. In most algorithms, this task can becomes computational bottleneck that takes up to 90\% of computation time~\cite{reggiani2002experimental}.

For algorithms that work with polyhedral models, \emph{Bounding Volume Hierarchies (BVHs)} are widely used to accelerate the tasks. The idea is to wrap polyhedral models with \emph{bounding volumes (BVs)} that allow faster overlap or separation distance test. Commonly used BVs are spheres, \emph{axis-aligned bounding boxes (AABB)}, \emph{oriented bounding boxes (OBB)} or \emph{discrete oriented polytopes ($k$-DOP)}. By splitting the models and wrap it recursively, this will generate a tree data structure called BVHs. To check overlap or separation distance between two models, we have to simultaneously traverse BVHs of the models (see Algorithm~\ref{alg:BVHTraversal}).

Despite its performance, BVHs also have drawback, which is the additional memory needed to describe the data structure, which usually consists of BVs descriptor (See Figure \ref{tbl:bv-memory}), pointers to its parent and children. Complex BVs usually could fit better into models however will require more memory to describe BVs. Due to the nature of BVHs traversal algorithms, memory access could becomes a bottleneck. To overcome this, some cache friendly BVHs layouts have been proposed e.g., cache-oblivious layout of BVHs (COLBVH) \cite{yoon2006cache}, and van Emde Boas layout \cite{van1975preserving}. Beside utilizing cache, reducing memory footprint could actually lead to better performance \cite{viitanen2017fast}.

In this paper, we proposed and implemented a novel compressed bounding volume hierarchies with the use case in collision detection \& proximity query. The  main  idea  is  to  compress bounding  volume  (BV)  descriptors  and  cluster  BVHs  into  a smaller  parts  (“treelet”)  that  fit  into  CPU  cache  while  at  the same time maintain random-access and automatic cache-aware data structure layouts. We decided to built our work on top of algorithm that make use of SIMD Instruction Sets to parallelize simultaneous traversal \cite{tan2019simdop}. Beside promising performance, the algorithm require additional memory footprint for storing BV descriptor in SIMD variables, which will be a good candidate to apply our compressed BVHs layout.

\begin{table}[h]
\label{table_example}
\begin{center}
\begin{tabular}{|c||c||c|}
\hline
BVs & Descriptor & Memory Use\\
\hline
Sphree & center + radius & 2 floats\\
AABB & min + max & 6 floats\\
k-DOP & number of k & k floats \\
\hline
\end{tabular}
\end{center}
\caption{BVs Memory Usage}
\label{tbl:bv-memory}
\end{table}

\begin{myalgorithm}
\If{$a$ and $b$ are both leaves}
{
checkPrimitives($a$, $b$)\;
}
\ElseIf{$a$ is leaf}
{
\ForAll{children $b_i$ of b}
{
\If{$a$ and $b_i$ intersect}
{
BVHtraversal($a$, $b_i$)\;
}
}
}
\ElseIf{$b$ is leaf}
{
\ForAll{children $a_i$ of a}
{
\If{$a_i$ and $b$ intersect}
{
BVHtraversal($a_i$, $b$)\;
}
}
}
\Else
{
\ForAll{children $a_i$ of a and $b_i$ of b}
{
\If{$a_i$ and $b_i$ intersect}
{
BVHtraversal($a_i$, $b_i$)\;
}
}
}
\caption{BVHtraversal( BV $a$, BV $b$ )}
\label{alg:BVHTraversal}
\end{myalgorithm}

\section{Previous Work}

In motion planning, many efficient algorithms that make use of different BVs have been proposed to accelerate collision test and proximity query, e.g., spheres \cite{hubbard1996approximating}, AABBs \cite{bergen1997efficient}, OBBs \cite{gottschalk1996obbtree}, a generalization of AABBs, BoxTree \cite{zachmann2002minimal}, memory optimized AABBs, k-DOPs \cite{klosowski1998efficient, zachmann1998rapid},  or convex hull trees \cite{ehmann2001accurate}. There also exist approach that combine several BVs into a three-stage sequence of BVs namely AABB, Sphere, and OBBs \cite{ferguson2008detection} or adding time as additional dimension into AABB-based BVHs \cite{schwesinger2015fast}. A recent algorithm make use of SIMD Instruction Sets inside CPU to accelerate the task by parallelize BVHs traversal algorithm and at the same time study the influence of branching factors \& splitting strategy in BVHs \cite{tan2019simdop}.

However, most algorithms did not take memory footprint into consideration. Actually, in the literature, there exist research that show an improvement in performance by reducing memory footprints \cite{kim2009racbvhs}.

In application like ray tracing, reducing BVHs memory footprint are being considered to avoid out of core technique especially when rendering huge scene either by reducing BVHs precision \cite{vaidyanathan2016watertight}, compress leaf with multiple polygons \cite{benthin2018compressed}, represent mesh using alternative representation \cite{lauterbach2007ray}, or implicit BVHs representation \cite{bauszat2010minimal} .

\section{Our Compressed Data Structure}
To maintain random-access and automatic cache-aware data structure layouts, we propose a novel compressed BVHs that have compression and decompresion method based on following components:
\subsection{Compression of BVs descriptor}
\label{subsec:compression-bv}
We decided to reduce precision of BVs descriptor by using half precison floating point, which only require 2 bytes instead of 4 bytes with single precision as reported in \cite{koskela2015using}. This instantly reducing our memory footprint by half and can be done using SIMD Instruction Sets $\_mm512\_cvtps\_ph$. Since SIMD does not allow direct operation for half float, we need to convert BVs descriptor back during overlap or separation distance test using $\_mm512\_cvtph\_ps$.

\subsection{Cluster BVHs into a smaller parts (“treelet”)}
We proposed of clustering BVHs into a smaller parts based on CPU cache size. 
Additionally, we use predictors based on the delta to the parent nodes.

\section{Results}
We have implemented our compressed BVHs layout using C++ and \emph{Intel Intrinsics functions} using Visual Studio 2019. We focused our implementation on the most recent AVX512 instruction sets. 

\begin{figure}
 \begin{subfigure}{.24\textwidth}
   \centering
 	\includegraphics[width=.58\textwidth]{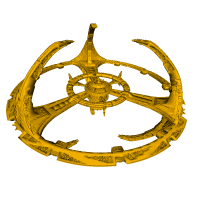}
 	\caption{\label{fig:ds9}}
 \end{subfigure}
 \begin{subfigure}{.24\textwidth}
   \centering
 	\includegraphics[width=0.58\textwidth]{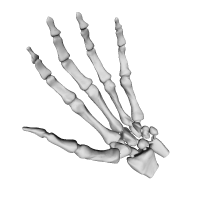}
 	\caption{\label{fig:hand}}
 \end{subfigure}
 \vspace{10px}
 
 \begin{subfigure}{.15\textwidth}
   \centering
 	\includegraphics[width=.50\textwidth]{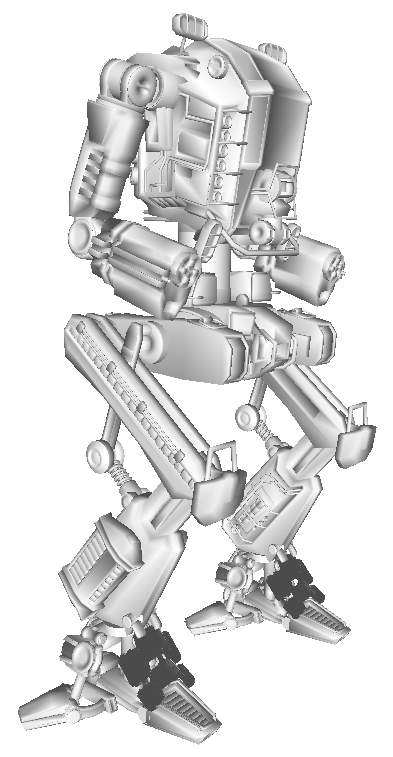}
 	\caption{\label{fig:robot-atst}}
 \end{subfigure}
 \begin{subfigure}{.15\textwidth}
   \centering
 	\includegraphics[width=1.3\textwidth]{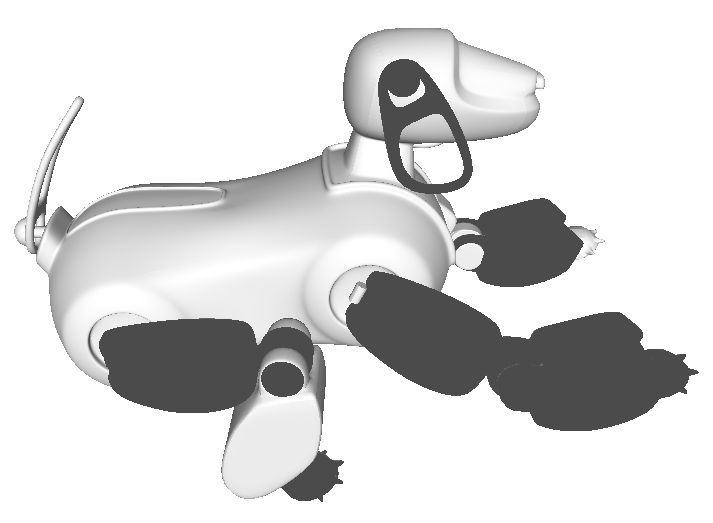}
 	\caption{\label{fig:robot-dog}}
 \end{subfigure}
 \begin{subfigure}{.15\textwidth}
   \centering
 	\includegraphics[width=0.50\textwidth]{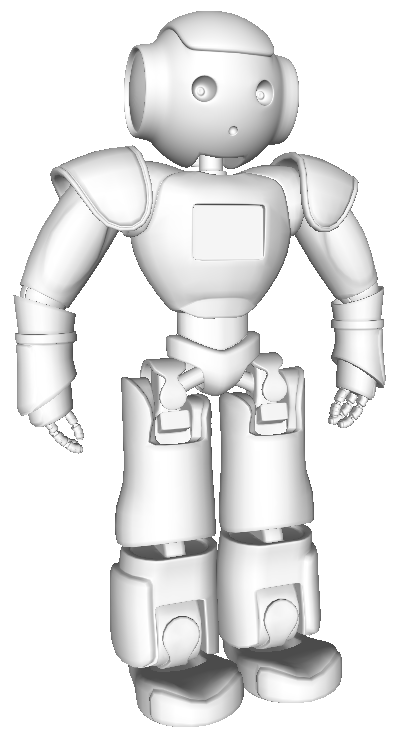}
 	\caption{\label{fig:robot-nao}}
 \end{subfigure}
 \vspace{10px}
 
 \caption{The objects we used in our timings: (a) ds9, (b) hand, (c) robot ATST, (d) robot dog, and (e) robot nao
 }
\label{fig:objects}
\end{figure}

To compare performance of our proposed BVHs layout, we use open benchmark for collision detection \& proximity query proposed by Tan et al. \cite{tan2020opencollbench}. Figure \ref{fig:objects} shows some of the used models with different shapes and resolutions in our timings: in particular, a ds9 station and a hand. We present all results in this section for the most time consuming distance preset, i.e. a distance of zero.

\begin{figure}
 \begin{subfigure}{.48\textwidth}
   \centering
 	\includegraphics[width=.95\textwidth]{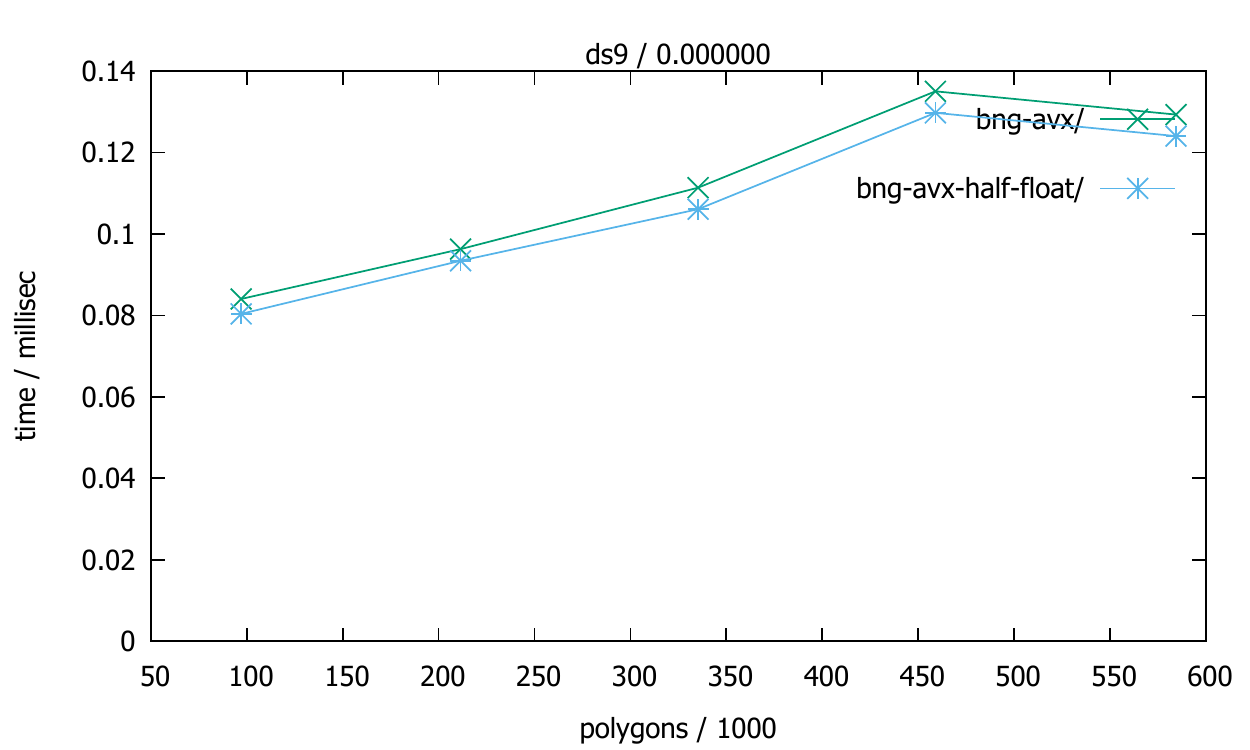}
 	\caption{\label{fig:ds9-half-float}}
 \end{subfigure}
 \vspace{10px}
 
 \begin{subfigure}{.48\textwidth}
   \centering
 	\includegraphics[width=.95\textwidth]{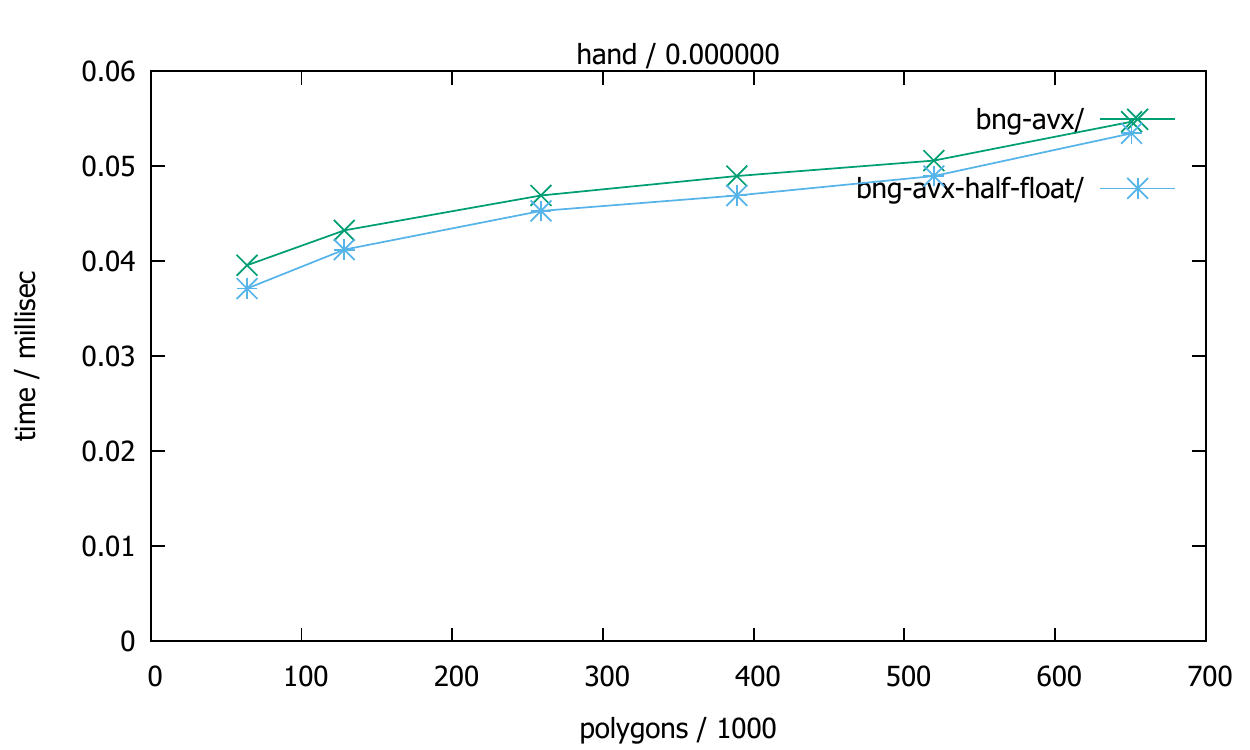}
 	\caption{\label{fig:hand-half-float}}
 \end{subfigure}
 \vspace{10px}
 
 \caption{BVHs simultaneous traversal timing for task collision detection using object (a) ds9 station, (b) hand , and (c) robot atst with half float to describe BVs.
 }
\label{fig:half-float}
\end{figure}

First, we evaluated the performance of using half precision floating point as described in subsection \ref{subsec:compression-bv}. Even though an additional operation needed to convert half precision to single precision before testing BV, the performance itself is actually better by around 5-10\% depends on objects used as shown by Figure \ref{fig:half-float}. This is due to the reduced memory traffic between CPU and RAM.

\section{Conclusion and Future Work}

\addtolength{\textheight}{-12cm}   







We have presented a novel compressed BVHs layout for collision detection \& proximity query. The  main  idea  is  to  compress bounding  volume  (BV)  descriptors  and  cluster  BVH  into  a smaller  parts  (“treelet”)  using predictor-corrector scheme with the predictor at a specific node in the BVH based on the chain of BVs upwards.

Our approach also opens up several directions for future work. For instance, we would like apply our layout into another BVs e.g., OBB or using another approach for compression, e.g., by quantify BV as integer and compress it using integer compression method \cite{lemire2018stream}. 



Finally, probably other applications using BVHs like ray tracing or occlusion computations could benefit from our compressed BVHs layout, too.

\bibliographystyle{IEEEtran} 
\bibliography{IEEEabrv,bibliography}

\end{document}